\documentclass[12pt,a4paper]{article}
\usepackage{epsf,exscale,times,graphicx,indentfirst,titlesec,caption,amssymb,amsmath,amstext,psfrag}

\titleformat{\section}{\normalsize\bfseries}{\thesection.}{0.3em}{}
\titleformat{\subsection}{\normalsize\bfseries}{\thesubsection.}{0.3em}{}

\begin{document}
\pagestyle{empty}
\begin{flushright}
\includegraphics[width=3.36cm]{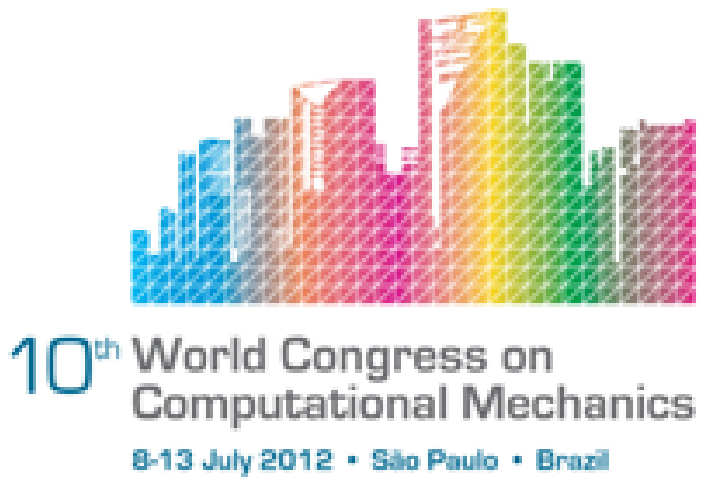}\\
\vskip0.5cm
\end{flushright}
%
%
\begin{flushleft}
{\fontsize{12}{16}\bf ON THE PHENOMENOLOGICAL MODELLING OF YIELD
SURFACE DISTORTION}\end{flushleft}
%
%
\begin{flushleft}
A. V. Shutov$^1$, J. Ihlemann$^1$\\
\end{flushleft}

\noindent $^1$ Department of Solid Mechanics,
Chemnitz University of Technology \\ (alexey.shutov@mb.tu-chemnitz.de)\\

\vspace{.5cm}
%
%
{\setlength{\baselineskip}{16pt} \noindent \textbf{Abstract.} {\it
A simple phenomenological approach to metal plasticity, including
the description of the strain-induced plastic anisotropy, is considered.
The advocated approach is exemplified by a two-dimensional
rheological analogy. This analogy provides insight into modelling of
nonlinear kinematic hardening of Armstrong-Frederick type combined with
a nonlinear distortional hardening.
In the previous publications of the authors, an interpolation rule between
the undistorted yield surface of a virgin material and the saturated yield
surface of a pre-deformed material was considered. In the current publication,
a somewhat more flexible approach is considered. Given a set of
convex symmetric key surfaces which correspond to different hardening stages,
the form of the yield surface is smoothly interpolated between these key surfaces.
Thus, any experimentally observed sequence of
symmetric convex yield surfaces can be rendered.
In particular, an arbitrary sharpening of the yield locus in
the loading direction combined with a flattening on
the opposite side can be taken into account. Moreover,
the yield locus evolves smoothly and its convexity is ensured at
each hardening stage.}
\vspace{.5cm}

\section{INTRODUCTION}

It is well known that already very small
plastic deformations may lead to a significant
change of the yield surface compared to
the initial state [3,19,8,9].
Such nonlinear effects like the residual stresses, springback, damage evolution,
and failure are highly dependent
on the accumulated
plastic anisotropy of the material.
Therefore, the proper description of the accumulated plastic anisotropy
is a challenging task.
Moreover, since the normality flow rule is implemented in most of phenomenological models,
the form of the yield surface has a significant impact on the overall material response
under non-proportional loading conditions.
For that reason, we concentrate on the phenomenological modelling of
plastic anisotropy with especial emphasis on
the distortional hardening.

Some of the recently developed models of distortional hardening
can be found, among others, in
[7, 1, 4,16,11,2,15].
Probably, the most simple approach to the distortional hardening
is based on the use of 2nd-rank tensors. Within this approach,
backstress-like tensors (directors) are introduced [12].
The applicability of different
constitutive assumptions was analyzed in [20] concerning
the description of the distorted yield surface.
The recent development of this approach is
presented in [6,13,5,17,18].

In the previous publication of authors [18],
an interpolation rule between the initial undistorted
yield surface and the saturated one was considered.
This saturated yield surface (which typically exhibits the maximum distortion)
was considered to be a material property.
The novelty of the current study as compared to
previous publication is as follows.
Now, a sequence of yield surfaces corresponding to different hardening stages can be
prescribed. For instance, such key
surfaces can be the initial undistorted one, some intermediate
yield surfaces, and the saturated one.
The interpolation between these key surfaces
retains the convexity and smoothness of the yield locus.

In order to provide an insight into the constitutive modelling, a two-dimensional
rheological model is considered in the current study, just as it was done in [18].
The translation and distortion of the yield surface as
well as its rotation depending on the recent loading path are
captured by the rheological model in a vivid way.
The construction of the closed system of constitutive equations,
basing on the rheological model, is straight-forward
in the case of small strains [18].
In particular, the kinematics of the material model, the
assumption for the energy storage and for the yield
function are motivated by the rheological model.
A strict proof of thermodynamic consistency can be provided, if
the yield surface remains convex and the
normality flow rule is considered, cf. [18].
Moreover, if finite strain plasticity is considered,
a system of constitutive equations can be obtained
using the elegant technique of Lion [10],
which is based on the consideration of rheological analogies.
As it was shown in [17], a similar technique
can be successfully implemented using two-dimensional rheological models, as well.

\section{TWO-DIMENSIONAL RHEOLOGICAL MODEL OF DISTORTIONAL HARDENING}

The idea to use two-dimensional rheological models
to motivate the constitutive equations of plasticity/viscoplasticity
with combined kinematic and distortional hardening was considered in [17].
A refined two-dimensional rheological model was presented later in [18].
Following [18], let us consider a mechanical system which consists of a
tank filled with a viscous fluid, a heavy solid
which rests on the flat bottom ($m.StV$), three elastic springs ($H_{\text{ext}}$,
$H_{\text{kin}}$, and $H_{\text{dis}}$) connected to the solid,
and two spheres ($m.N_{\text{kin}}$ and $m.N_{\text{dis}}$) floating on the surface
of the fluid (Fig. \ref{fig1}).\footnote{An animated version of the
rheological model with only one modified Newton element is available
at http://www.youtube.com/watch?v=QEPc3pixbC0}

\begin{figure}[h]\centering
\psfrag{A}[m][][1][0]{$A$}
\psfrag{SV}[m][][1][0]{$m.StV$}
\psfrag{H1}[m][][1][0]{$H_{\text{ext}}$}
\psfrag{H2}[m][][1][0]{$H_{\text{dis}}$}
\psfrag{H3}[m][][1][0]{$H_{\text{kin}}$}
\psfrag{N1}[m][][1][0]{$m.N_{\text{dis}}$}
\psfrag{N2}[m][][1][0]{$m.N_{\text{kin}}$}
\psfrag{T}[m][][1][0]{$\theta$}
\psfrag{AA}[m][][1][0]{a)}
\psfrag{BB}[m][][1][0]{b)}
\scalebox{0.9}{\includegraphics{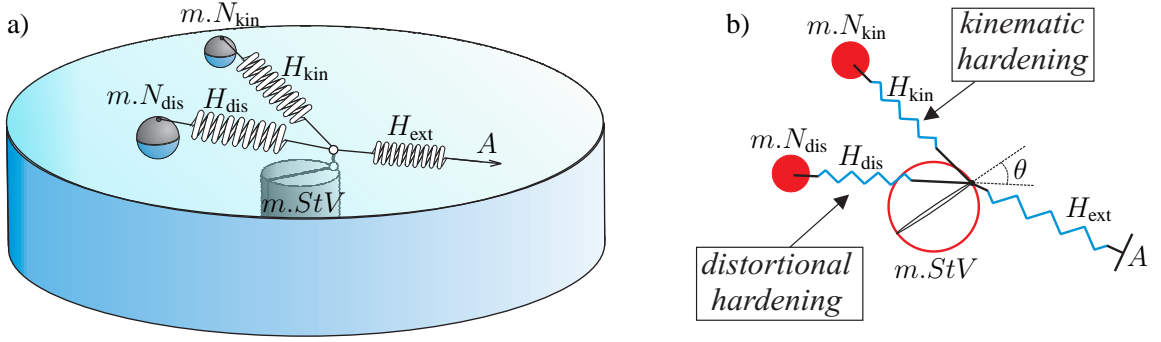}}
\caption{a) Two-dimensional rheological model of combined kinematic
and distortional hardening. The model
is built up of a modified St.-Venant element ($m.StV$),
Hooke-bodies $H_{\text{ext}}$, $H_{\text{kin}}$, $H_{\text{dis}}$, and
modified Newton elements $m.N_{\text{kin}}$ and $m.N_{\text{dis}}$;
b) Rheological model seen from above. The
angle between the ($m.StV$)-axis and $H_{\text{dis}}$ is
denoted by $\theta$. The critical friction force of the
($m.StV$)-element depends on $\theta$.
  \label{fig1}}
\end{figure}

\begin{figure}[h]\centering
\psfrag{SV}[m][][1][0]{$m.StV$}
\psfrag{S}[m][][1][0]{$\vec{\sigma}_{H}$}
\psfrag{F}[m][][1][0]{$\vec{\sigma}_{N}$}
\psfrag{G}[m][][1][0]{$\vec{\sigma}$}
\psfrag{SM}[m][][1][0]{$-\vec{\sigma}_{H}$}
\psfrag{A}[m][][1][0]{$A$}
\psfrag{B}[m][][1][0]{$B$}
\psfrag{E}[m][][1][0]{$\vec{\varepsilon}_{H}$}
\psfrag{V}[m][][1][0]{$\frac{d}{d p}\vec{\varepsilon}_{N}$}
\psfrag{H}[m][][1][0]{$H$}
\psfrag{N}[m][][1][0]{$m.N$}
\psfrag{T}[m][][1][0]{$\theta$}
\psfrag{SE}[m][][1][0]{$\vec{\sigma}_{\text{eff}}$}
\psfrag{XK}[m][][1][0]{$-\vec{x}_{\text{k}}$}
\psfrag{XD}[m][][1][0]{$-\vec{x}_{\text{d}}$}
\psfrag{AA}[m][][1][0]{a)}
\psfrag{BB}[m][][1][0]{b)}
\psfrag{CC}[m][][1][0]{c)}
\scalebox{0.9}{\includegraphics{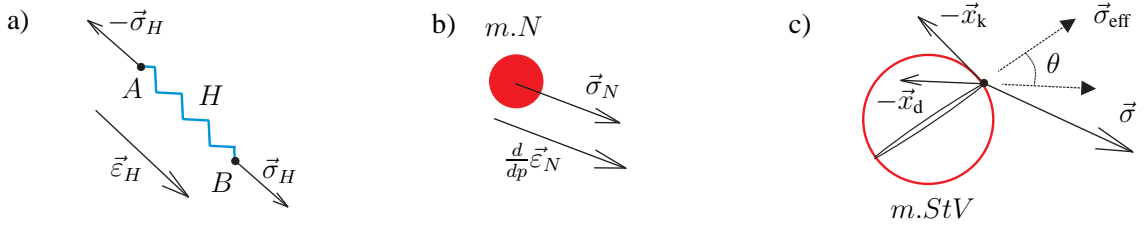}}
\caption{Behavior of idealized two-dimensional bodies: a) Hooke-body $H$;
b) Modified Newton-body $m.N$;
c) Modified St.-Venant element $m.StV$. For this element, the friction depends
on the angle $\theta$. \label{fig2}}
\end{figure}

Let us recall the rheological properties of the
idealized bodies (for more details, the reader is referred to [17,18]):
\begin{itemize}
\begin{item} ($H$): For the Hooke-bodies (see Fig. \ref{fig2}a),
we put $\vec{\sigma}_{H} = c \ \vec{\varepsilon}_{H}$, where
$\vec{\sigma}_{H}$ stands for the spring force,
$\vec{\varepsilon}_{H} = \overrightarrow{AB} \in \mathbb{R}^2$ is the elongation of the spring,
and $c \geq 0$ is a fixed stiffness.
\end{item}
\begin{item} ($m.N$): The two-dimensional Newton element is represented by
a sphere which is floating on the surface.
We assume that
the fluid resistance $\vec{\sigma}_{N}$ to the motion of the sphere is proportional to
its velocity $\frac{d}{d t}\vec{\varepsilon}_{N}$. Thus,
$\frac{d}{d t}\vec{\varepsilon}_{N} = \varkappa \ \vec{\sigma}_{N}$, where
$\varkappa \geq 0$ is a fixed viscosity parameter.
Next, in order to obtain
rate-independent constitutive equations,
the physical time $t$ is formally replaced
by the accumulated inelastic arc-length (Odqvist parameter) $p$.
Thus, we get for the modified Newton element (($m.N$)-element)
(see Fig. \ref{fig2}b):
\begin{equation}\label{NewtPos}
\frac{d}{d p}\vec{\varepsilon}_{N} = \varkappa \ \vec{\sigma}_{N}.
\end{equation}
\end{item}
\begin{item}
($m.StV$):
The heavy solid rests on the bottom of the tank and there is a
friction between them. By $\vec{\sigma}$, $-\vec{x}_{\text{k}}$, and $-\vec{x}_{\text{d}}$
denote now the forces acting on this solid due to the elongation of
the Hooke-bodies $H_{\text{ext}}$, $H_{\text{kin}}$, and $H_{\text{dis}}$, respectively  (see Fig. \ref{fig2}c).
The force $\vec{\sigma}$ will be understood as an external load;
$\vec{x}_{\text{k}}$ and $\vec{x}_{\text{d}}$ will be responsible for the effects similar to
kinematic and distortional hardening, respectively.
The effective force acting on the solid is thus given by
$\vec{\sigma}_{\text{eff}} = \vec{\sigma} - \vec{x}_{\text{k}} -
\vec{x}_{\text{d}}$.
Let the axis of the ($m.StV$)-element be always oriented along
$\vec{\sigma}_{\text{eff}}$.
The ($m.StV$)-element remains at rest as long as $\| \vec{\sigma}_{\text{eff}} \| \leq \sqrt{2/3} K$, where
$\sqrt{2/3} K >0$ is a nonconstant friction.
More precisely: Let $\theta$ be the angle between the ($m.StV$)-axis and $\vec{x}_{\text{d}}$:
$\theta = \text{arccos}\Big( \frac{\vec{\sigma}_{\text{eff}} \ \cdot \ \vec{x}_{\text{d}}}
{\|\vec{\sigma}_{\text{eff}}\| \ \|\vec{x}_{\text{d}}\|}\Big)$.
Moreover, let $\alpha = \|\vec{x}_{\text{d}}\| / x_d^{\text{max}}$ be a distortion parameter, which
is a unique function of $\|\vec{x}_{\text{d}}\|$.
Here, $x_d^{\text{max}} >0$ is the upper bound for $\|\vec{x}_{\text{d}}\|$,
therefore we get $\alpha \in [0, 1]$. Finally, we consider the friction to be a function of $\theta$ and $\alpha$:
$K = \bar{K}(\theta, \alpha) \ K_0$, where $K_0 >0$ is a fixed parameter.
A suitable ansatz for $\bar{K}(\theta, \alpha)$ is essential for the proper description
of distortional hardening. One simple geometric ansatz was considered in [18]. An
alternative rule will be presented in the next section.
\end{item}
\end{itemize}

\section{CONSTRUCTION OF DIRECTION-DEPENDENT YIELD STRESS}

In this section, a flexible rule for the computation of
the friction function $\bar{K}(\theta, \alpha)$ is suggested.
Note that this friction function can be interpreted
as a non-dimensional yield stress (cf. [18]).
Unlike the ansatz presented previously in [18], this
rule is based on the interpolation between certain key surfaces.
In particular, such key surfaces can be identified experimentally at different
hardening stages (cf. Section 4).

\begin{figure}[h]\centering
\psfrag{D}[m][][1][0]{$\text{El}_{key}^{1}$}
\psfrag{H}[m][][1][0]{$(\alpha = 0)$}
\psfrag{F}[m][][1][0]{$\text{El}_{key}^{i}$}
\psfrag{M}[m][][1][0]{$(0 < \alpha < 1)$}
\psfrag{E}[m][][1][0]{$\text{El}_{key}^{M}$}
\psfrag{Z}[m][][1][0]{$(\alpha = 1)$}
\psfrag{K}[m][][1][0]{key surfaces}
\psfrag{EL}[m][][1][0]{$\text{El}(\bar{K}( \cdot , \alpha))$}
\psfrag{U}[m][][1][0]{$\vec{e}_1$}
\psfrag{T}[m][][1][0]{$\theta$}
\psfrag{X}[m][][1][0]{$\vec{y}$}
\psfrag{x}[m][][1][0]{$\vec{y}$}
\psfrag{O}[m][][1][0]{$\vec{0}$}
\psfrag{AA}[m][][1][0]{a)}
\psfrag{BB}[m][][1][0]{b)}
\scalebox{0.8}{\includegraphics{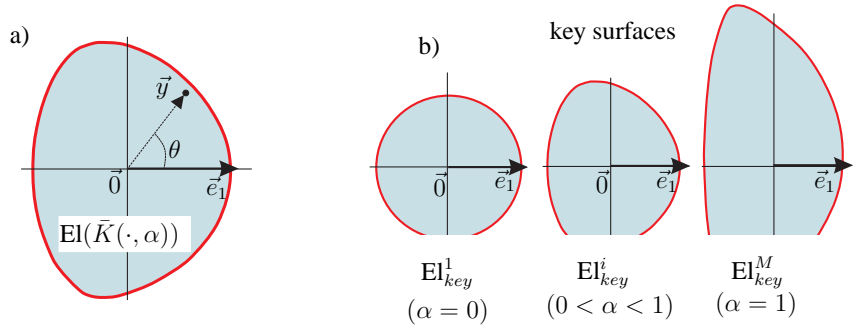}}
\caption{a) Interpretation of the function $\bar{K}(\cdot, \alpha)$ in
terms of corresponding convex set $\text{El}(\bar{K}( \cdot , \alpha))$;
b) Example of three key surfaces.
Each key surface must be smooth and convex. The first key surface
corresponds to zero distortion (surface with $\alpha=0$).
The last surface describes the fully saturated distortion (surface with $\alpha=1$). \label{fig3}}
\end{figure}

Let $\vec{e}_1 = (1,0) \in \mathbb{R}^2$.
Similar to [18], for each distortion parameter $\alpha \in [0,1]$ we consider the set $\text{El}(\bar{K}( \cdot , \alpha))$ which
consists of $\vec{y} \in \mathbb{R}^2$ such that $\| \vec{y} \| \leq  \bar{K}(\theta, \alpha)$ (see Fig. \ref{fig3}a). Here,
$\theta \in [0, \pi]$ is the angle between $\vec{y}$ and $\vec{e}_1$.
In particular, the upper half of the boundary of this set is described by
\begin{equation}\label{Boundparam}
 \vec{y}(\theta, \alpha) = (y_1(\theta, \alpha), y_2(\theta, \alpha)) = \bar{K}(\theta, \alpha) (\cos(\theta), \sin(\theta)), \quad
 \theta \in [0,\pi].
\end{equation}

Let us consider $M$ different convex symmetric subsets $\text{El}_{key}^{i}, \ i=1,...,M$, as shown in Fig. \ref{fig3}b.
Each of these sets is related to a certain hardening stage, and each stage is characterized by the
corresponding value of the distortion parameter $\alpha^{i}_{key}, \ i=1,...,M$:
\begin{equation}\label{subdivision1}
    0=\alpha^{1}_{key} < \alpha^{2}_{key} < ... < \alpha^{M}_{key}=1.
\end{equation}
Assume that a unit disc corresponds to the initial (undistorted) stage ($\alpha=0$).

Our goal now is to construct a continuous function $\bar{K}(\theta, \alpha)$, such that:
\begin{itemize}
\item[i:] \ $\bar{K}(\theta, \alpha) > 0$ \ for all $\theta \in [0, \pi], \ \alpha \in [0,1]$,
\item[ii:] \ $\bar{K}(0, \alpha) = 1 $ \ for all $\alpha \in [0,1]$,
\item[iii:] \ $\frac{\displaystyle \partial \bar{K}(\theta, \alpha)}{\displaystyle \partial \theta}|_{\theta=0} =
\frac{\displaystyle \partial \bar{K}(\theta, \alpha)}{\displaystyle \partial \theta}|_{\theta=\pi} = 0 $ \ for all $\alpha \in [0,1]$,
\item[iv:] \ the set $\text{El}(\bar{K}( \cdot , \alpha))$ is convex for all $\alpha \in [0,1]$,
\item[v:] \ the boundary of $\text{El}(\bar{K}( \cdot , \alpha))$ is smooth for all $\alpha \in [0,1]$,
\item[vi:] the set $\text{El}(\bar{K}( \cdot , \alpha^{i}_{key}))$ coincides with $\text{El}_{key}^{i}$ for all $i = 1,...,M$.
\end{itemize}

Let $\vec{n}^{\alpha}(\theta)$ be the outward unit normal to the boundary of $\text{El}(\bar{K}( \cdot , \alpha))$ at the
point $\vec{y}(\theta, \alpha)$.
The angle between this normal and $\vec{e}_1$ will be denoted by $Y^{\alpha}(\theta)$ (see Fig. \ref{fig4}a).
According to (ii), $\bar{K}(0, \alpha)$ is given.
Thus, as it will be shown in the following, the function $\bar{K}(\cdot, \alpha)$ is uniquely
determined by $Y^{\alpha}(\cdot)$, for each fixed $\alpha \in [0,1]$.

\begin{figure}[h]\centering
\psfrag{EL}[m][][1][0]{$\text{El}(\bar{K}( \cdot , \alpha))$}
\psfrag{U}[m][][1][0]{$\vec{e}_1$}
\psfrag{T}[m][][1][0]{$\theta$}
\psfrag{N}[m][][1][0]{$\vec{n}^{\alpha}(\theta)$}
\psfrag{Y}[m][][1][0]{$Y^{\alpha}(\theta)$}
\psfrag{P}[m][][1][0]{$Y^{\alpha}(\theta) - \theta$}
\psfrag{X}[m][][1][0]{$\vec{y}(\theta, \alpha)$}
\psfrag{Z}[m][][1][0]{$\vec{y}(\theta + \Delta \theta, \alpha)$}
\psfrag{F}[m][][1][0]{$\Delta \theta$}
\psfrag{O}[m][][1][0]{$\vec{0}$}
\psfrag{AA}[m][][1][0]{a)}
\psfrag{BB}[m][][1][0]{b)}
\scalebox{0.8}{\includegraphics{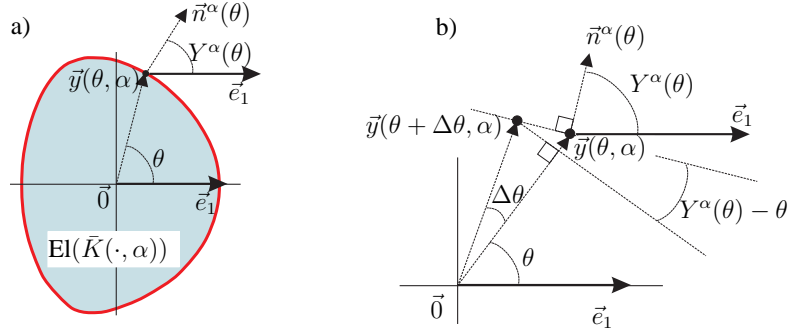}}
\caption{a) Definition of the outward normal $\vec{n}^{\alpha}(\theta)$ and the angle $Y^{\alpha}(\theta)$.
b) Geometric considerations behind equation \eqref{DiffEq}. Here, $\Delta \theta$ is an infinitesimal perturbation of $\theta$.
The corresponding change in $\bar{K}(\theta, \alpha) = \|\vec{y}(\theta, \alpha)\|$ equals
$\Delta  \bar{K}(\theta, \alpha) = \bar{K}(\theta, \alpha) \tan(\theta - Y^{\alpha}(\theta)) \Delta \theta$. \label{fig4}}
\end{figure}

For each of the subsets  $\text{El}_{key}^{i}$ ($i=1,...,M$), let us consider the corresponding
functions $Y^{i}_{key}(\theta)$.
In other words, the functions $Y^{i}_{key}(\theta)$ describe the direction of the normality vector
corresponding to the key surfaces.
Since $\text{El}_{key}^{i}$ are convex with a smooth boundary,
the functions $Y^{i}_{key}(\theta)$ are continuous, monotonically increasing, $Y^{i}_{key}(0)=0$,
$Y^{i}_{key}(\pi)=\pi$, and $| Y^{i}_{key}(\theta) - \theta | \leq \pi/2$ \ for all $\theta \in [0, \pi]$.
In what follows it will be shown that the problem (i)-(vi) can be reduced to an equivalent one: Find $Y^{\alpha}(\theta)$
such that:
\begin{itemize}
\item[I:] \ $ | Y^{\alpha}(\theta) - \theta | < \pi/2$ \ for all $\theta \in [0, \pi], \ \alpha \in [0,1]$,
\item[III:] \  $Y^{\alpha}(0) = 0 $, $Y^{\alpha}(\pi) = \pi $ \ for all $\alpha \in [0,1]$,
\item[IV:] \ $Y^{\alpha}(\theta)$ is a monotonically increasing function of $\theta$, for each fixed $\alpha \in [0,1]$,
\item[V:] \ $Y^{\alpha}(\theta)$ is a continuous function of $\theta$, for each fixed $\alpha \in [0,1]$,
\item[VI:]  $Y^{\alpha^{i}_{key}}(\theta) = Y^{i}_{key}(\theta)$ for all $i = 1,...,M$, $\theta \in [0, \pi]$.
\end{itemize}

In the current study, we use the following rule to construct $Y^{\alpha}(\theta)$ which satisfies (I)-(VI).
We interpolate between two key surfaces: For $\alpha \in [\alpha^{i}_{key}, \alpha^{i+1}_{key}]$ we put
\begin{equation}\label{InterpolationY}
Y^{\alpha}(\theta) = \frac{\alpha^{i+1}_{key} - \alpha}{ \alpha^{i+1}_{key} - \alpha^{i}_{key}} Y^{i}_{key}(\theta) +
 \frac{\alpha - \alpha^{i}_{key}}{ \alpha^{i+1}_{key} - \alpha^{i}_{key}}  Y^{i+1}_{key}(\theta), \quad
 \text{for all} \ \theta \in [0, \pi].
\end{equation}
Now it remains to restore $\bar{K}(\cdot, \alpha)$ from $Y^{\alpha}(\cdot)$.
Some geometric considerations (see Fig. \ref{fig4}b) yield the following differential equation:
\begin{equation}\label{DiffEq}
\frac{\displaystyle \partial \bar{K}(\theta, \alpha)}{\displaystyle \partial \theta} =
\bar{K}(\theta, \alpha) \tan(\theta - Y^{\alpha}(\theta)), \ \text{for} \ \theta \in [0, \pi]; \quad
\bar{K}(0, \alpha) =1.
\end{equation}
There exists a simple closed-form solution of problem \eqref{DiffEq}, if
$Y^{\alpha}(\theta)$ is a piecewise linear function of $\theta$.
Indeed, let us consider a subdivision
\begin{equation}\label{subdivision2}
    0=\theta_{1} < \theta_{2} < ... < \theta_{N}=\pi.
\end{equation}
Suppose that the functions $Y^{i}_{key}(\theta)$ are linear within
$[\theta_j, \theta_{j+1}]$.
Thus, for $\theta \in [\theta_j, \theta_{j+1}]$, we get
\begin{equation}\label{InterpOfY}
Y^{\alpha}(\theta) =
\frac{\displaystyle \theta_{j+1} - \theta}{\displaystyle \theta_{j+1} - \theta_j} Y_{j}^{\alpha} +
\frac{\displaystyle \theta - \theta_j}{\displaystyle \theta_{j+1} - \theta_j} Y_{j+1}^{\alpha},
\end{equation}
where the values $Y_{j}^{\alpha} = Y^{\alpha}(\theta_j)$ and $Y_{j+1}^{\alpha} = Y^{\alpha}(\theta_{j+1})$
are obtained using the interpolation rule \eqref{InterpolationY}.

Considering $\alpha$ as a fixed parameter, we get from \eqref{DiffEq}
\begin{equation}\label{RestoreK}
 d \big(\ln (\bar{K}(\theta, \alpha)) \big) = \tan(\theta - Y^{\alpha}(\theta)) d \theta.
\end{equation}
Next, we abbreviate
\begin{equation}\label{RestoreK2}
k:= 1 + \frac{\displaystyle Y_{j}^{\alpha} -
Y_{j+1}^{\alpha}}{\displaystyle \theta_{j+1} - \theta_{j}}, \quad
l:=\frac{\displaystyle \theta_{j} \displaystyle
Y_{j+1}^{\alpha} - \theta_{j+1} Y_{j}^{\alpha}}{\displaystyle \theta_{j+1} - \theta_{j}}.
\end{equation}
Substituting \eqref{InterpOfY} into \eqref{RestoreK},
one gets for
$\theta \in [\theta_j, \theta_{j+1}]$ 
\begin{equation}\label{RestoreK3}
 d \big(\ln (\bar{K}(\theta, \alpha)) \big) = \tan(k \theta + l) d \theta.
\end{equation}
Integrating \eqref{RestoreK3} from $\theta_j$ to $\theta$, we get for all $\theta \in [\theta_j, \theta_{j+1}]$
\begin{equation}\label{RestoreK4}
\ln (\bar{K}(\theta, \alpha)) =
- \frac{1}{k} \ln \Big|\frac{\cos{k \theta + l}}{\cos{k \theta_j + l}}\Big| +
\ln (\bar{K}(\theta_j, \alpha))  \quad  \text{for} \ k \neq 0,
\end{equation}
\begin{equation}\label{RestoreK5}
\ln (\bar{K}(\theta, \alpha)) =
\tan(l) (\theta - \theta_j) + \ln (\bar{K}(\theta_j, \alpha))  \quad  \text{for} \ k = 0.
\end{equation}
Taking into account that $\ln (\bar{K}(0, \alpha)) = 0$, $\ln (\bar{K}(\theta, \alpha))$ can be computed
through sequential evaluation of $\ln (\bar{K}(\cdot, \alpha))$ for
all $\theta_j \leq \theta$.

\textbf{Remark 1} Let us show that the inequality (I) implies (i). Indeed, if (I) holds, then
$ \tan(\theta - Y^{\alpha}(\theta)) > - \infty $. Combining this with \eqref{RestoreK}, one gets
$\ln (\bar{K}(\theta, \alpha)) > - \infty$, or, equivalently,  $\bar{K}(\theta, \alpha) > 0$.

\textbf{Remark 2} Observe that the outward normal is explicitly given by
$\vec{n}^{\alpha}(\theta) = \\ (\cos(Y^{\alpha}(\theta)),\sin(Y^{\alpha}(\theta)))$.
This expression can be utilized by constructing an appropriate flow rule (cf. equation (43) in [18] for the normality flow rule).

\textbf{Remark 3} In this study we assume that the key surfaces are given. Indeed, it was shown in
[18] that the form of the key surface can be directly associated to the form of the corresponding yield surface
in $(\sigma_1, \sqrt{3} \sigma_{12})$-space. Moreover, if some yield surfaces in $(\sigma_1, \sigma_2)$-space
are available, they also can be used after an appropriate affine transformation.

\section{EXAMPLE}

Let us consider the experimental data on the evolution of the yield surface under
uniaxial tension, presented by Phillips and Tang in [14].
The progressing plastic anisotropy in
commercially pure aluminum 1100-0 at room temperature is shown in Fig. \ref{fig5}a.
As it can be seen, the plastic deformation involves
isotropic softening, kinematic translation of the yield surface, and its distortion.\footnote{The
reduction of the elastic range \emph{under proportional loading conditions} is referred to as isotropic softening.}
The yield surfaces, which are considered as key surfaces, are depicted in Fig. \ref{fig5}b.
Here, we put $\alpha^1_{key} =0$, $\alpha^2_{key} =0.5$, $\alpha^3_{key} =1$.\footnote{In general, $\alpha^2_{key}$ should be considered as a material parameter.}
Note that the size of the elastic domain for all key surfaces is the same along $\vec{e}_1 = (1,0)$, and
the origin $(0,0)$ is located exactly in the middle.
It is done so in order to decouple the yield surface distortion from the kinematic and isotropic hardening.
The corresponding functions $Y^{\alpha}(\theta)$ are plotted in Fig.\ref{fig6}a against $\theta$.
These functions are monotonically increasing and continuous.
The interpolation results are shown in Fig. \ref{fig6}b for different values of the distortion
parameter $\alpha$. As it can be seen, all surfaces remain convex and smooth.

\begin{figure}\centering
\psfrag{C}[m][][1][0]{$\alpha = \alpha^1_{key} = 0$}
\psfrag{D}[m][][1][0]{$\alpha = \alpha^2_{key} = 0.5$}
\psfrag{F}[m][][1][0]{$\alpha = \alpha^3_{key} = 1$}
\psfrag{AA}[m][][1][0]{a)}
\psfrag{BB}[m][][1][0]{b)}
\scalebox{0.8}{\includegraphics{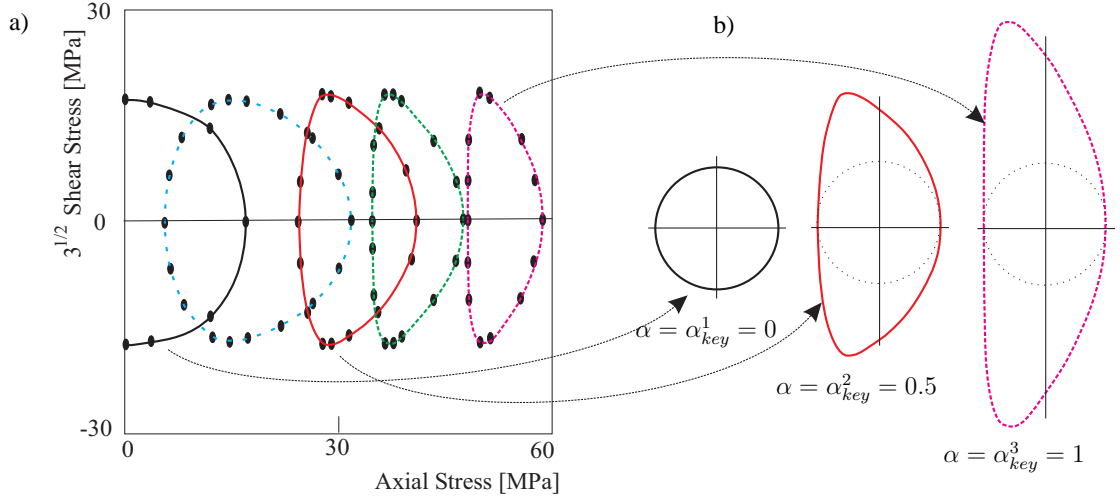}}
\caption{a) Experimental data for commercially pure aluminum 1100-0 at room temperature [14];
b) The three key surfaces. The form of the key surfaces coincides with the form of
the experimentally measured yield surfaces.\label{fig5}}
\end{figure}

\begin{figure}\centering
\psfrag{A}[m][][1][0]{$\alpha   = 0$}
\psfrag{B}[m][][1][0]{$\alpha = 0.5$}
\psfrag{D}[m][][1][0]{$\alpha  = 1$}
\psfrag{F}[m][][1][0]{a)}
\psfrag{G}[m][][1][0]{b)}
\psfrag{T}[m][][1][0]{$\theta$}
\psfrag{Y}[m][][1][0]{$Y^{\alpha}(\theta)$}
\psfrag{N}[m][][1][0]{$0$}
\psfrag{P}[m][][1][0]{$\pi$}
\scalebox{0.75}{\includegraphics{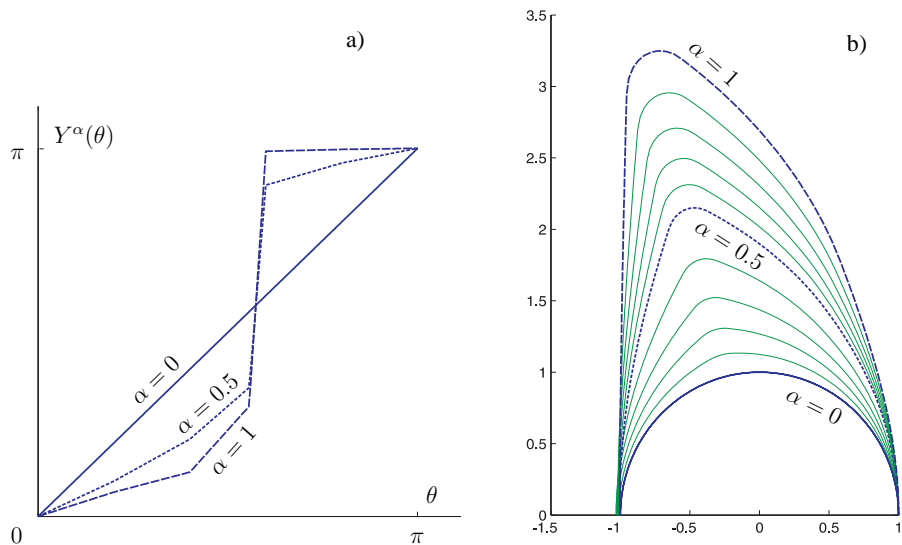}}
\caption{a) Plots of $Y^{\alpha}(\theta)$ for different values of $\alpha$;
b) Interpolation results for $\alpha =0, 0.1, 0.2, ..., 1$. Only the upper half is shown.\label{fig6}}
\end{figure}

\section{CONCLUSION}

In this work, the main emphasis is made on the phenomenological description of the yield surface distortion.
Toward that end, a practical construction of the function $\bar{K}(\theta, \alpha)$ is considered.
In terms of the two-dimensional rheological interpretation, this function represents a direction-dependent
friction of the modified St.-Venant element ($m.StV$). On the other hand, in terms of a plasticity model, this
function plays a role of the yield stress. The approach advocated in this study utilizes an
interpolation between certain key surfaces (boundaries of $\text{El}_{key}^{i}, \ i=1,...,M$).
In particular, it can be guaranteed that the resulting surfaces
are convex and smooth, such that the outward normal is
well defined. The proposed interpolation rule is simple and easy to implement.
Using this approach, any experimentally
observed sequence of symmetric convex yield surfaces
can be rendered with desired accuracy. Such property is
especially important, if the normality flow rule is used as a
constitutive assumption.

\section*{Acknowledgements}

\noindent The financial support provided by DFG within SFB 692 is acknowledged.
\section{REFERENCES}
\noindent{
\begin{tabular}{lp{150mm}}
\hspace{-3mm}{[1]}&
Dafalias Y.F., Schick D., Tsakmakis C., ``A simple model for describing
yield surface evolution''. \emph{Lecture note in applied and computational
mechanics, K. Hutter and H. Baaser, eds., Springer, Berlin.} 169-201, 2002.
\\[1mm]
\hspace{-3mm}{[2]}&
Dafalias Y.F., Feigenbaum H.P., ``Directional distortional hardening
in plasticity within thermodynamics''. \emph{Recent Advances in Mechanics}
61-78, 2011.
\\[1mm]
\hspace{-3mm}{[3]}&
Dannemeyer S., ``Zur Ver\"anderung der Flie\ss fl\"ache von Baustahl bei
mehrachsiger plastischer Wechselbeanspruchung''. \emph{Braunschweig (Carolo-
Wilhelmina University)} 1999.
\\[1mm]
\hspace{-3mm}{[4]}&
Feigenbaum H.P., Dafalias Y.F., ``Directional distortional hardening
in metal plasticity within thermodynamics''. \emph{International Journal of Solids
and Structures} 44, 7526-7542, 2007.
\\[1mm]
\hspace{-3mm}{[5]}&
Feigenbaum H.P., Dafalias Y.F., ``Simple model for directional distortional
hardening in metal plasticity within thermodynamics''. \emph{Journal of
Engineering Mechanics} 134 9, 730-738, 2008.
\\[1mm]
\hspace{-3mm}{[6]}&
Fran\c{c}ois M., ``A plasticity model with yield surface distortion for non
proportional loading''. \emph{International Journal of Plasticity} 17, 703-717, 2001.
\\[1mm]
\hspace{-3mm}{[7]}&
Grewolls G., Krei\ss ig R.,  ``Anisotropic hardening -- numerical application
of a cubic yield theory and consideration of variable r-values for sheet
metal''. \emph{European Journal of Mechanics A/Solids} 20, 585-599, 2001.
\\[1mm]
\end{tabular}}
\begin{tabular}{lp{150mm}}
\hspace{-3mm}{[8]}&
Khan A.S., Pandey A., Stoughton T.,  ``Evolution of subsequent yield
surfaces and elastic constants with finite plastic deformation. Part II: A
very high work hardening aluminum alloy (annealed 1100 Al)''. \emph{International
Journal of Plasticity} 26, 1421-1431, 2010.
\\[1mm]
\hspace{-3mm}{[9]}&
Khan A.S., Pandey A., Stoughton T., ``Evolution of subsequent yield
surfaces and elastic constants with finite plastic deformation. Part III:
Yield surface in tension-tension stress space (Al 6061T 6511 and annealed
1100 Al)''. \emph{International Journal of Plasticity} 26, 1432-1441, 2010.
\\[1mm]
\hspace{-3mm}{[10]}&
Lion A., ``Constitutive modelling in finite thermoviscoplasticity: a physical
approach based on nonlinear rheological elements''. \emph{International Journal
of Plasticity} 16, 469-494, 2000.
\\[1mm]
\hspace{-3mm}{[11]}&
Noman M., Clausmeyer T., Barthel C., Svendsen B., Hu´etink J., Riel,
M., ``Experimental characterization and modeling of the hardening
behavior of the sheet steel LH800''. \emph{Materials Science and Engineering A}
527, 2515-2526, 2010.
\\[1mm]
\hspace{-3mm}{[12]}&
Ortiz M., Popov E.P., ``Distortional hardening rules for metal plasticity''.
\emph{J. Engng Mech.} 109, 1042-1058, 1983.
\\[1mm]
\hspace{-3mm}{[13]}&
Panhans S., Krei\ss ig R., ``A viscoplastic material model of overstress
type with a non-quadratic yield function''. \emph{European Journal of Mechanics
A/Solids} 25, 283-298, 2006.
\\[1mm]
\hspace{-3mm}{[14]}&
Phillips A., Tang J.-L., ``The effect of loading path on the yield surface at
elevated temperatures''. \emph{Int. J. Solids Structures.} 8, 463-474, 1972.
\\[1mm]
\hspace{-3mm}{[15]}&
Pietryga M.P., Vladimirov I.N., Reese S., ``A finite deformation model
for evolving flow anisotropy with distortional hardening including experimental
validation''. \emph{Mechanics of Materials} 44, 163-173, 2012.
\\[1mm]
\hspace{-3mm}{[16]}&
Plesek J., Feigenbaum H.P., Dafalias Y.F., ``Convexity of yield surface
with directional distortional hardening rules''. \emph{Journal of Engineering
Mechanics} 136 4, 477-484, 2010.
\\[1mm]
\hspace{-3mm}{[17]}&
Shutov A.V., Panhans S., Krei\ss ig R., ``A phenomenological model
of finite strain viscoplasticity with distortional hardening''. \emph{ZAMM} 91 8,
653-680, 2011.
\\[1mm]
\hspace{-3mm}{[18]}&
Shutov A.V., Ihlemann J., ``A viscoplasticity model with an enhanced control of the
yield surface distortion''. \emph{arXiv-preprint} http://arxiv.org/abs/1204.0086  2012.
\\[1mm]
\hspace{-3mm}{[19]}&
Steck E., Ritter R., Peil U., Ziegenbein A., ``Plasticity of Materials: Experiments,
Models''. \emph{Deutsche Forschungsgemeinschaft, Computation}
(Wiley-VCH Verlag GmbH), 2001.
\\[1mm]
\hspace{-3mm}{[20]}&
Wegener K., Schlegel M., ``Suitability of yield functions for the approximation
of subsequent yild surfaces''. \emph{International Journal of Plasticity} 12, 1151-1177, 1996.
\\[1mm]
\end{tabular}}
\end{document}